\documentclass[aps,pre,showpacs,twocolumn]{revtex4}

\usepackage{amsmath}
\usepackage{amssymb}
\usepackage{amsfonts}
\usepackage{mathrsfs}
\usepackage{graphicx}

\begin{document}

\title{Optimal target search on a fast folding polymer chain with volume
exchange}

\author{Michael A. Lomholt}
\author{Tobias Ambj{\"o}rnsson}
\author{Ralf Metzler}
\affiliation{NORDITA, Blegdamsvej 17, 2100 Copenhagen \O, Denmark}

\begin{abstract}
We study the search process of a target on a rapidly folding polymer (`DNA')
by an ensemble of particles (`proteins'), whose search combines 1D diffusion
along the chain, L{\'e}vy type diffusion mediated by chain looping, and
volume exchange. A rich behavior of the search process is obtained with
respect to the physical parameters, in particular, for the optimal search.
\end{abstract}


\maketitle

\textbf{\emph{Introduction.}} L{\'e}vy flights (LFs) are random walks whose
jump lengths $x$ are distributed like $\lambda(x)\simeq|x|^{-1-\alpha}$ with
exponent $0<\alpha<2$ \cite{hughes}. Their probability density
to be at position $x$ at time $t$ has the characteristic function
$P(q,t)\equiv\int_{-\infty}^{\infty}e^{iqx}P(x,t)dx=\exp\left(-D_{\mathrm{L}}
|q|^{\alpha}t\right)$, a consequence of the generalized central limit theorem
\cite{levy}; in that sense, LFs are a natural extension of normal Gaussian
diffusion ($\alpha=2$). LFs occur in a wide range of systems
\cite{report}; in particular, they represent an optimal search
mechanism in contrast to locally oversampling Gaussian search
\cite{stanley}. Dynamically, LFs can be described by a space-fractional
diffusion equation $\partial P/\partial t=D_{\mathrm{L}}\partial^{\alpha}
P(x,t)/\partial|x|^{ \alpha}$, a convenient basis to introduce additional
terms, as shown below. $D_{\mathrm{L}}$ is a diffusion constant of
dimension $\mathrm{cm}^{\alpha}/\mathrm{sec}$, and the fractional derivative is
defined via its Fourier transform, $\mathscr{F}\{\partial^\alpha P(x,t)
/\partial |x|^\alpha\}=-|q|^\alpha P(q,t)$ \cite{report}. LFs exhibit
superdiffusion in the sense that $\langle|x|^{\zeta}\rangle^{2/\zeta}\simeq
(D_{\mathrm{L}}t)^{2/\alpha}$ ($0<\zeta<\alpha$), spreading
faster than the linear dependence of standard diffusion ($\alpha=2$).

\begin{figure}
\includegraphics[width=4.8cm]{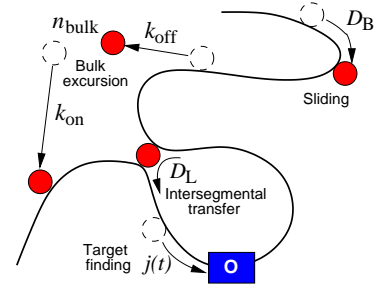}
\caption{Search mechanisms in Eq.~(\ref{eq:themodel}).}
\label{fig:ownmechanisms}
\end{figure}

A prime example of an LF is linear particle diffusion to next neighbor
sites on a fast folding (`annealed') polymer that permits intersegmental
jumps at chain contact points (see Fig.~\ref{fig:ownmechanisms}) due to
polymer looping \cite{igor,dirk}. The contour length $|x|$
stored in a loop between such contact points is distributed in 3D like 
$\lambda(x)\simeq|x|^{-1-\alpha}$, where $\alpha=1/2$ for Gaussian chains
($\theta$ solvent), and $\alpha\approx 1.2$ for self-avoiding walk chains
(good solvent) \cite{duplantier}.

While non-specifically bound \cite{audun}, proteins can diffusively slide
along the DNA backbone in search of their specific target site,
as long as the binding energy does not exceed a certain limit \cite{slutsky}.
Under overstretching conditions preventing looping, pure 1D sliding search
could be observed in vitro \cite{mark}. In absence of the stretching force,
the combination of intersegmental jumps (LF component) and 1D sliding may be
a good approximation to the motion of binding proteins or enzymes along a DNA.
In general,
however, proteins detach to the volume and, after a bulk excursion, reattach
successively before reaching the target. This mediation by de- and
(re)adsorption rates $k_{\mathrm{off}}$ and $k_{\mathrm{on}}$ is described by
the Berg-von Hippel model sketched in Fig.~\ref{fig:ownmechanisms} \cite{bvh}.
We here explore by combination of analytical and numerical analysis for the
first time (1) the combination of 1D sliding, intersegmental transfer
\emph{and} volume exchange, (2) a particle number \emph{density} instead of a
single searching protein; and (3) the explicit determination of the first
arrival to the target, per se a non-trivial problem for LFs
\cite{chechkinbvp}. Note that, although the process we study is a
generic soft matter problem, we here adopt the DNA-protein language for
illustration.

\textbf{\emph{Theoretical description.}} In our description of the target
search process, we use the density per length $n(x,t)$ of proteins on
the DNA as the relevant dynamical quantity ($x$ is the distance along
the DNA contour). Apart from intersegmental transfer, we include 1D
sliding along
the DNA with diffusion constant $D_{\mathrm{B}}$, protein dissociation
with rate $k_{\mathrm{off}}$ and (re)adsorbtion with rate $k_{\mathrm{on}}$
from a bath of proteins of concentration $n_{\mathrm{bulk}}$.
The dynamics of $n(x,t)$ is thus governed by the equation \cite{REM2}
\begin{eqnarray}
\nonumber
\frac{\partial}{\partial t}n(x,t)=&&\left(D_{\rm B} \frac{\partial^2}{\partial
x^2}+D_{\rm L} \frac{\partial^\alpha}{\partial |x|^\alpha}-k_{\rm
off}\right)n(x,t)\nonumber\\
&&+k_{\rm on}n_{\rm bulk}-j(t)\delta(x).
\label{eq:themodel}
\end{eqnarray}
Here, $j(t)$ is the flux into the target located at $x=0$. We determine
the flux $j(t)$ by assuming that the target is perfectly absorbing: $n(
0,t)=0$. Be initially the system at equilibrium, except that the target is
unoccupied; then, the initial protein density is $n_0=n(x,0)=k_{\rm on}
n_{\rm bulk}/k_{\rm off}$ \cite{REM}. The total number
of particles that have arrived at the target up to time $t$ is
$J(t)=\int_0^t d t'\;j(t')$. We derive explicit analytic
expressions for $J(t)$ in different limiting regimes, and study
the general case numerically. We use $J(t)$ to obtain
the mean first arrival time $T$ to the target; in
particular, to find the value of $k_{\rm off}$ that minimizes $T$.

\begin{figure}
\includegraphics[width=8.4cm]{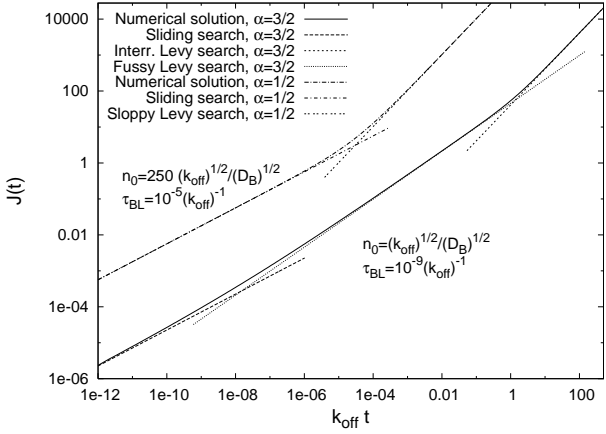}
\caption{Number of proteins arrived at the target up to $t$.
Numerical solutions of Eq.~(\ref{eq:inteq}) and limiting regimes.}
\label{fig:JIeq}
\end{figure}

To proceed, we Laplace and Fourier transform Eq.~(\ref{eq:themodel}):
\begin{eqnarray}
\nonumber
u n(q,u)-2\pi n_0 \delta(q)=&-\left(D_{\rm B} q^2+D_{\rm L}
|q|^\alpha+k_{\rm off}\right)\\
&\hspace*{-2cm}\times
n(q,u)+2\pi k_{\rm on}n_{\rm bulk}\delta(q)/u-j(u),
\label{eq:Laplace}
\end{eqnarray}
with $n(q,u)=\mathscr{L}\{n(q,t)\}$. Integration over $q$
produces $J(u)=j(u)/u=n_0/\left[u^2W_0(u)\right]$ due to
the perfect absorption condition $n(0,u)=(2\pi)^{-1}\int dq\,
n(q,u)=0$. Or,
\begin{equation}
\label{eq:inteq}\textstyle\int_0^t d t'\;W_0(t-t')J(t') = n_0 t
\end{equation}
in the $t$-domain. Eq.~(\ref{eq:inteq}) is a Volterra integral equation of
the first kind, whose kernel $W_0$ is read off Eq.~(\ref{eq:Laplace}):
\begin{equation}
\label{eq:W0u}W_0(u)=\int_{-\infty}^\infty\frac{dq}{2\pi}
\frac{1}{D_{\rm B} q^2+D_{\rm L} |q|^\alpha+k_{\rm off}+u},
\end{equation}
that is the Laplace transform of the Green's function of $n(x,t)$ at $x=0$.
Back-transforming, we obtain
$W_0(t)=(2\pi)^{-1}\int_{-\infty}^\infty dq\exp\left(-(D_{\rm B} q^2+D_{\rm L}
|q|^\alpha+k_{\rm off})t\right)$,
which has a singularity at $t=0$.
Eq.~(\ref{eq:inteq}) can be solved numerically by approximating
$J(t)$ by a piecewise linear function, converting the integral
equation to a linear set of equations. Typical plots are shown in
Fig.~\ref{fig:JIeq}.

Eq.~(\ref{eq:W0u}) reveals only two relevant time scales: $k_{\rm off}^{-1}$
and $\tau_{\rm BL}=(D_{\rm B}^ \alpha/D_{\rm L}^2)^{1/(2-\alpha)}$. We
now obtain asymptotic results for small and large $(k_{\rm off}+u)$,
compared to $\tau_{\rm BL}^{-1}$.

{\it $k_{\rm off}+u\gg \tau_{\rm BL}^{-1}$}: In this limit, the
denominator of the integrand in Eq.~(\ref{eq:W0u}) is dominated
either by the term $D_{\rm B}q^2$ or by $k_{\rm off}+u$ for any
$q$; we find the approximation \cite{thesymbol}
\begin{equation}\label{eq:W0ularge}
W_0(u)\sim \left.W_0(u)\right|_{D_{\rm L}=0}=\left[D_{\rm B}
(k_{\rm off}+u)\right]^{-1/2}/2\;.
\end{equation}

{\it $k_{\rm off}+u\ll\tau_{\rm BL}^{-1}$ and $\alpha{\it >1}$ (`connected
LFs')}: Here, a singularity exists at small $q$ as $k_{\rm
off}+u\to 0$. For finite but small $k_{\rm off}+u\to 0$, the integrand
is dominated by the $D_{\rm L}|q|^\alpha$ term compared to $D_{\rm
B}q^2$ at small $q$, yielding
\begin{equation}\label{eq:W0usmall}
W_0(u)\sim \left[\alpha \sin(\pi/\alpha) D_{\rm L}^{1/\alpha}(k_{\rm
off}+u)^{1-1/\alpha} \right]^{-1}.
\end{equation}

{\it $k_{\rm off}+u\ll\tau_{\rm BL}^{-1}$ and $\alpha {\it <1}$ (`disconnected
LFs')}: Now, the singularity is weak, and the integral becomes
\begin{equation}\label{eq:W0usmallless}
W_0(u)\sim \left[(2-\alpha)\sin\left(\frac{1-\alpha}{2-\alpha}\pi\right) \sqrt{D_{\rm B}\tau_{\rm BL}^{-1}}
\right]^{-1}.
\end{equation}

From these limits, we now infer the behavior of $J(t)$, based on Tauberian
theorems stating that $J(t)$ at $t\to 0$ is determined by $J(u)$ at $u\to
\infty$, and vice versa \cite{hughes}. We discover a rich variety of
domains, compare Tab.~\ref{tab:Jregimes}:

{\it (1.) Sliding search}: Desorption from the DNA can be neglected for
times $t\ll k_{\rm off}^{-1}$. If also $t\ll\tau_{\rm BL}$,
Eq.~(\ref{eq:W0ularge}) with $k_{\rm off}=0$ by inverse Laplace transform
leads to
\begin{equation}\label{eq:smallt}
J(t)\sim\left(t/\tau_1\right)^{\gamma_1},\,\gamma_1=1/2,\,
\tau_1=\pi/(16 D_B n_0^{1/\gamma_1}).
\end{equation}
In this regime, only the 1D sliding mechanism matters.

{\it (2.) Fussy L{\'e}vy search}: For $\tau_{\rm
  BL}\ll t\ll k_{\rm off}^{-1}$ ($\alpha>1$), the LF dominates the
flux into the target; from Eq.~(\ref{eq:W0usmall}),
\begin{equation}\label{eq:intert}
J(t)\sim(t/\tau_2)^{\gamma_2},\,\gamma_2=1/\alpha,\,
\tau_2=C_2/(D_L n_0^{1/\gamma_2}),
\end{equation}
where $C_2=\{\Gamma (1+1/\alpha)/[\alpha \sin (\pi/\alpha)]\}^{\alpha}$. Now,
LFs are the overall dominating mechanism. This contrasts:

{\it (3.) Sloppy L{\'e}vy search}: For $\alpha<1$, $t\gg\tau_{\rm BL}$,
and $k_{\rm off}^{-1}\gg\tau_{\rm BL}$, we obtain from
Eq.~(\ref{eq:W0usmallless})
\begin{equation}\label{eq:largetB}
J(t)\sim\left( \frac{t}{\tau_3}\right)^{\gamma_3},\, \gamma_3=1,\,
\tau_3=C_3\frac{D_B^{\alpha/[2(2-\alpha)]-1/2}}{D_L^{1/(2-\alpha)}
n_0^{1/\gamma_3}},
\end{equation}
and $C_3=\{(2-\alpha)\sin([1-\alpha]\pi/[2-\alpha])\}^{-1}$. For $\alpha<1$,
even the step length $\int d x\,|x|\lambda(x)$ diverges, making it impossible
for the protein to hit a small target solely by LF, and local sampling by 1D
sliding becomes vital. At longer times, volume exchange mediated by
$k_{\rm off}$ enters:

{\it (4.) Interrupted L{\'e}vy search}: For $\alpha>1$ and $t\gg
k_{\rm off}^{-1}\gg \tau_{\rm BL}$ we can ignore $u$ in
Eq.~(\ref{eq:W0usmall}), yielding
\begin{equation}\label{eq:larget}
J(t)\sim\left(t/\tau_4\right)^{\gamma_4},\,\gamma_4=1,
\tau_4=C_4/(D_L^{1/\alpha}k_{\rm off}^{1-1/\alpha} n_0^{1/\gamma_4}),
\end{equation}
with $C_4=1/[\alpha\sin(\pi/\alpha)]$.
The search on the DNA is dominated by LFs, interrupted
by 3D volume excursions.

{\it (5.) Interrupted sliding search}: If $\tau_{\rm BL}\gg k_{\rm
off}^{-1}$, LFs will not contribute at any $t$.
Instead we find from Eq.~(\ref{eq:W0ularge})
\begin{equation}\label{eq:larget2}
J(t)\sim\left(t/\tau_5\right)^{\gamma_5},\, \gamma_5=1,\,
\tau_5=1/(2D_B^{1/2}k_{\rm off}^{1/2}n_0^{1/\gamma_5})
\end{equation}
for $t\gg k_{\rm off}^{-1}$.
This is sliding-dominated search with 3D excursions. There exist
three scaling regimes for $1<\alpha<2$, and two for $0<\alpha<1$;
see Fig.~\ref{fig:JIeq} and Tab.~\ref{tab:Jregimes}.

We found that the relevant time scales $k_{\rm off}^{-1}$ and $\tau_{\rm BL}$
together with $\alpha$ give rise to 5 basic search regimes, each
characterized by an exponent $\gamma_i$ and characteristic time scale
$\tau_i$. In particular, we saw that $J(t)\sim (t/\tau_i)^{\gamma_i}$,
where the exponent $\gamma_i\neq 1$ for the first two regimes ($i=1,2$);
in the other cases, we have $\gamma_i=1$. The stable index $\alpha$
characterizing the polymer statistics thus strongly influences the overall
search. Also note that $J(t)\simeq t$ when $t\gg k_{\rm off}^{-1}$, or
$t\gg\tau_{\rm BL}$ and $\alpha<1$. The characteristic time scales $\tau_i$,
since $J(t)\simeq n_0$, scale like $\tau_i\simeq n_0^{-1/\gamma_i}$. As any
integral $I=\int_0^\infty dt f(t/\tau_i)$ can be transformed by $s\equiv
t/\tau_i$ to $I=\tau_i \int_0^\infty ds f(s)$, it is $I\simeq \tau_i$. Thus,
we find that the mean first arrival time scales like $T=\tau_i \int_0^\infty ds
\exp(-s^{\gamma_i})=\tau_i\Gamma(1/\gamma_i)/\gamma_i\simeq n_0^{-1/\gamma_i}$
(see below) whenever a single of the five regimes dominates the integral.
In particular, the variation of $T^{-1}$ with the line density $n_0$ ranges
from quadratic (1D sliding) over $n_0^{\alpha}$ in the fussy L{\'e}vy regime
($1<\alpha<2$) to linear,
the latter being shared by sloppy L{\'e}vy and bulk mediated search.
Note that if 1D sliding is the sole prevalent mechanism, we recover the result
$T=\pi/[8D_{\mathrm{B}} n_0^2]$ of Ref.~\cite{mark}.

\begin{table}
\begin{tabular}{|l|c|c|c|}
\hline Regime & $0<\alpha<1$ & $1<\alpha<2$ & $J\sim(t/\tau_i)^{\gamma_i}$\\
\hline $t\ll\{\tau_{\rm BL},k_{\rm off}^{-1}\}$ & Sliding & Sliding &
$\gamma_1=1/2$\\
$\tau_{\rm BL}\ll t\ll k_{\rm off}^{-1}$ & Sloppy L{\'e}vy & Fussy L{\'e}vy &
$\gamma_3=1\,|\,\gamma_2=\alpha^{-1}$\\
$\tau_{\rm BL}\ll  k_{\rm off}^{-1} \ll t$ & Sloppy L{\'e}vy & Int. L{\'e}vy &
$\gamma_3=1\,|\,\gamma_4=1$\\
$\{t,\tau_{\rm BL}\}\gg k_{\rm off}^{-1}$ & Int.~Sliding& Int.~Sliding&
$\gamma_5=1$ \\
\hline
\end{tabular}
\caption{Summary of search regimes. See text.}
\label{tab:Jregimes}
\end{table}

\textbf{\emph{Optimal search.}} We now address the optimal search of the target,
i.e., which $k_{\rm off}$ minimizes the mean first arrival time $T$ when
$D_{\rm B}$, $D_{\rm L}$, $k_{\rm on}$, the DNA length $L$, and the total
amount of proteins are fixed. To quantify the latter, we define $l_{\rm DNA}
\equiv L/V$, where $V$ is the system volume. The overall protein volume
density is then $n_{\rm total}=l_{\rm DNA}n_0+n_{\rm bulk}$.
With the equilibrium condition $k_{\rm off}n_0=k_{\rm
on}n_{\rm bulk}$, this yields $n_0=n_{\mathrm{total}}k_{\mathrm{on}}/
\left(k_{\mathrm{off}}+k_{\mathrm{on}}'\right)$ and a corresponding
expression for $n_{\mathrm{bulk}}$; here, $k_{\rm on}'=
k_{\rm on}l_{\rm DNA}$ is the inverse average time a single protein spends
in the bulk solvent before (re)binding to the DNA.

To extract the mean first arrival time $T$, we reason as follows (compare
Ref.~\cite{mark}): The total number of proteins that have arrived at the
target between $t'=0$ and $t$ is $J(t)$. If $N$ is the overall number of
proteins, the probability for an individual protein to have arrived at the
target is $J(t)/N$. In the limit of large $N$, we obtain the
survival probability of the target (no protein has arrived) as
\begin{equation}\label{eq:Psurv}
P_{\rm surv}(t)=\lim_{N\to
\infty}\left(1-J(t)/N\right)^N=\exp\left[-J(t)\right],
\end{equation}
and thus $T=\int_0^\infty d t\;P_{\rm surv}(t)$. Note that for LFs, the first
arrival is crucially different from the first passage \cite{chechkinbvp}.

The optimization is complicated by the exponential function
in Eq.~(\ref{eq:Psurv}). However, both in vitro and in vivo, $n_{\rm total}$
(and hence $n_0$) is in many cases sufficiently small, such that the
relevant regime is $J(t)\propto t$ (i.e., we can approximate $W_0(u)$ by
$W_0(u=0)$). The mean first arrival time in this linear regime becomes
\begin{equation}\label{eq:linearT}
T=W_0(u=0)[(k_{\rm off}+k_{\rm on}')/k_{\rm on}'][l_{\rm DNA}/n_{\rm
total}].
\end{equation}
We observe a tradeoff in the optimal value $k_{\rm off}^{\rm opt}$,
that minimizes $T$: The fraction $k_{\rm on}'/(k_{\rm off}+k_{\rm on}')$
of bound proteins shrinks with increasing $k_{\rm off}$,
increasing $T$. Counteracting is the decrease of $W_0(u=0)$ (and $T$) with
growing $k_{\rm off}$.

\begin{figure}
\includegraphics[width=6.8cm]{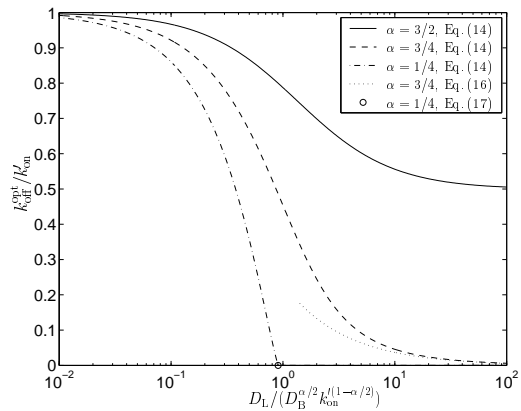}
\caption{Optimal choice of off rate $k_{\rm off}$ as function of the
LF diffusion constant, from numerical evaluation of Eq.~(\ref{eq:linearT}).
The circle on the abscissa marks where $k_{\rm off}^{\rm opt}$ becomes
0 (Eq.~(\ref{eq:onequarterreg})).}
\label{fig:threeopts}
\end{figure}

Numerical solutions to the optimal search are shown in
Fig.~\ref{fig:threeopts} for different $\alpha$.
Three different regimes emerge:

(i) Without LFs ($D_{\rm L}\to 0$ or $D_{\rm L}\ll D_{\rm B}^{
\alpha/2}(k_{\rm on}')^{1-\alpha/2}$),
from Eq.~(\ref{eq:W0ularge}) with $W_0$ at $u=0$, we obtain
$k_{\rm off}^{\rm opt}=k_{\rm on}'$:
the proteins should spend equal amounts of time in bulk and on the DNA.
This corresponds to the result obtained for single protein searching on
a long DNA \cite{slutsky,benichou}. Two additional regimes unfold for
strong LF search, $D_{\rm L}\to \infty$:

(ii) For $\alpha>1$, where Eq.~(\ref{eq:W0usmall}) applies, we find
\begin{equation}
k_{\rm off}^{\rm opt}\sim (\alpha-1)k_{\rm on}':
\end{equation}
The optimal off rate shrinks linearly with decreasing $\alpha$.

(iii) For $\alpha<1$, the value of $k_{\rm off}^{\rm opt}$ approaches
zero as $D_{\rm L}\to\infty$: The sloppy LF mechanism becomes so
efficient that bulk excursions become irrelevant. More precisely,
for $1/2<\alpha<1$ as $D_{\rm L}$ goes to infinity,
\begin{equation}
\label{eq:threequarterlimit}
k_{\rm off}^{\rm opt}\sim \left(\frac{
(2-\alpha)(1-\alpha)\sin\left(\frac{1-\alpha}{2-\alpha}\pi\right)
}{ \alpha^2\sin\left(\frac{2\alpha-1}{\alpha}\pi\right) }k_{\rm
on}'\tau_{\rm BL}^{1/\alpha-1}\right)^{\frac{\alpha}{2\alpha-1}}
\end{equation}
At $\alpha=1/2$, we observe a qualitative change:
When $\alpha<1/2$, the rate $k_{\rm off}^{\rm opt}$ reaches
zero for all \emph{finite} $D_{\rm L}$ satisfying
\begin{equation}
\label{eq:onequarterreg}
\tau_{\rm
BL}^{-1}\ge\frac{(1+\alpha)\sin\left([1-\alpha]\pi/[2-\alpha]
\right)}{(2-\alpha)\sin\left([1-2\alpha]\pi/[2-\alpha]\right)}k_{\rm
on}'\;.
\end{equation}
Note that when $\alpha<1$, the spread of the LF ($\simeq t^{1/\alpha}$) grows
faster than the number of sites visited ($\simeq t$), rendering the mixing
effect of bulk excursions insignificant. A scaling argument to understand
the crossover at $\alpha=1/2$ relates the probability density of first
arrival with the width ($\simeq t^{1/\alpha}$) of the Green's function of an
LF, $p_{\mathrm{fa}}\simeq t^{-1/\alpha}$. The associated mean
arrival time becomes finite for $0<\alpha<1/2$, even for the infinite chain
considered here.

\textbf{\emph{Discussion.}}
Eq.~(\ref{eq:themodel}) phrases the target search problem as
a fractional diffusion-reaction equation with point sink.
This formulation pays tribute to the fact that
for LFs, the first arrival differs from the first passage: With the long-tailed
$\lambda(x)$ of an LF, the particle can repeatedly jump across the target
without hitting, the first arrival becoming less efficient than the first
passage \cite{chechkinbvp}.

A borderline role is played by the Cauchy case
$\alpha=1$, separating connected (mean jump length $\langle|x|\rangle$ exists)
and disconnected LFs. For $\alpha<1$, the number
of visited sites grows slower than the width of the search region and the LF
mimics the uncorrelated jumps of bulk excursion; the latter becomes obsolete
for high LF diffusivity $D_{\mathrm{L}}$. Below $\alpha=1/2$, bulk excursions
already for finite $D_{\mathrm{L}}$ are undesirable. A similar observation
can be made for the scaling of the mean search time $T$ with the L{\'e}vy
diffusivity $D_{\rm L}$, that is proportional to the rate an LF is performed:
For $\alpha>1$ in the interrupted L{\'e}vy search, $T\simeq D_{\rm L}^{-1/
\alpha}$, whereas $T\simeq D_{\rm L}^{-1/(2-\alpha)}$ in the sloppy L{\'e}vy
search, where $\alpha<1$. The L{\'e}vy component is thus taken most profit
of when $\alpha$ approaches 1. Generally, too short jumps, leading to local
oversampling, as well as too long jumps, missing the target, are unfavorable.

A crucial assumption of the model, analogous to the derivation in
Ref.~\cite{dirk}, is that on the time scale of the diffusion process
the polymer chain appears annealed; otherwise, individual jumps are
no longer uncorrelated \cite{igor}. Generally, for proteins $D_{\mathrm{B}}$
is fairly low, and can be further lowered by adjusting the salt
condition, so that the conditions for the annealed case can be met.
Conversely, by increasing $D_{\mathrm{B}}$ in respect to
the polymer dynamics, leading to a higher probability to use the
same looping-induced `shortcut' repeatedly, it might be possible to investigate
the turnover from LF motion to `paradoxical diffusion' of the
quenched polymer case \cite{igor}.

Single molecule studies can probe the dynamics of the target search and the
quantitative predictions of our model \cite{mark,ctn}.
Monitoring the target finding dynamics may also be a novel way of
investigating soft matter properties regarding both polymer equilibrium
configurations, giving rise to $\alpha$, and its dynamics.
With respect to the first arrival properties, it would be interesting
to study the gradual change of the polymer properties from self-avoiding
behavior in a good solvent to Gaussian chain statistics under $\theta$
or dense conditions.


In a next step, it will be of interest to explore effects on
the DNA looping behavior due to (a) the occurrence of local
denaturation bubbles performing as hinges \cite{marko}, whose dynamics
can be understood from statistical approaches \cite{tobias}; or (b) kinks
imprinted on the DNA locally by binding proteins. In the presence of
different protein species, the first arrival method may provide a
way to probe protein crowding effects
to expand existing models toward the in vivo situation.

\textbf{\textit{Conclusion.}} Our search model reveals
rich behavior in dependence of the LF diffusivity $D_{\mathrm{L}}$ and exponent
$\alpha$. In particular, we found two crossovers for the optimal search that
we expect to be accessible experimentally.
In that sense, our model system is richer than the 2D albatross
search model \cite{stanley}. We note that in the Cauchy case $\alpha=1$
additional logarithmic contributions are superimposed to the power laws
\cite{REM3}. Moreover, long-time memory effects may occur in the process;
in the protein search, e.g., there are indications that both the
sliding search through stronger protein-DNA interactions \cite{slutsky}
and the volume diffusion through crowding effects are
sub\-dif\-fu\-si\-ve \cite{report}.

We thank I.~M. Sokolov and U.~Gerland for discussions.


\begin{thebibliography}{99}
\bibitem{REM2} For identical proteins, their mutual avoidance is actually
included in Eq.~(\ref{eq:themodel}), as on encounter it does not matter
whether they deflect each other or swap identities.
\bibitem{REM} Note that the dimension of the on and off rates differ; while
$[k_{\mathrm{off}}]=\mathrm{sec}^{-1}$, we chose $[k_{\mathrm{on}}]=
\mathrm{cm}^2/\mathrm{sec}$.
\bibitem{thesymbol}
The symbol $\sim$ implies that the relative difference
vanishes, e.g.:
$\lim_{k_{\rm off}+u\to\infty}\left.W_0(u)\right|_{D_{\rm L}=0}/W_0(u)=1$.
\bibitem{REM3} For $\alpha=1$ and $k_{\mathrm{off}}+u\ll
\tau_{\mathrm{BL}}^{-1}$ we find $W_0(u)\sim(\pi D_{\mathrm{L}})^{-1}\log\left\{
1/[\tau_{\mathrm{BL}}(u+k_{\mathrm{off}})]\right\}$.
At $k_{\mathrm{off}}^{-1}\gg t\gg\tau_{\mathrm{BL}}$,
Tauberian theorems \cite{hughes}
lead to the logarithmic correction
$J(t)\sim t/[\tau_6\log(t/\tau_{\mathrm{BL}})]$, with $\tau_6=1/(\pi n_0D_
{\mathrm{L}})$, while for $t\gg k_{\mathrm{off}}^{-1}\gg\tau_{\mathrm{BL}}$,
we have $J(t)\sim t/[\tau_6\log\{1/(\tau_{\mathrm{BL}}k_{\mathrm{off}})\}]$.
\end{thebibliography}
\end{document}